\documentclass[lnicst,sechang,a4paper]{svmultln}
\usepackage{amssymb}
\usepackage{graphicx}

\usepackage{url}
\urldef{\mailsa}\path|{mohsin.khan, valtteri.niemi}@helsinki.fi|
\usepackage[pdfpagelabels,hypertexnames=false,breaklinks=true,bookmarksopen=true,bookmarksopenlevel=2]{hyperref}

\usepackage{color}
\usepackage[numbers]{natbib}
\usepackage{calc}
\usepackage{siunitx}
\DeclareSIUnit\mt{\milli\tesla} 
\begin{document}

\mainmatter  

\title{AES and SNOW 3G are Feasible Choices for a 5G Phone from Energy Perspective}

\titlerunning{AES and SNOW 3G are Feasible Choices}

%
%
\author{Mohsin Khan%
\and Valtteri Niemi}  %

\institute{University of Helsinki, Department of Computer Science,\\
P.O. Box 68 (Gustaf H\"allstr\"omin katu 2b)\\
FI-00014 University of Helsinki\\
Finland\\
\mailsa
}

%
%

\toctitle{AES and SNOW 3G are Feasible Choices}
\tocauthor{Authors' Instructions}
\maketitle

\begin{abstract}
The aspirations for a 5th generation (5G) mobile network are high. It has a vision of unprecedented data-rate and extremely pervasive connectivity. To cater such aspirations in a mobile phone, many existing efficiency aspects of a mobile phone need to be reviewed. We look into the matter of required energy to encrypt and decrypt the huge amount of traffic that will leave from and enter into a 5G enabled mobile phone. In this paper, we present an account of the power consumption details of the efficient hardware implementations of AES and SNOW 3G. We also present an account of the power consumption details of LTE protocol stack on some cutting edge hardware platforms. Based on the aforementioned two accounts, we argue that the energy requirement for the current encryption systems AES and SNOW 3G will not impact the battery-life of a 5G enabled mobile phone by any significant proportion.
\end{abstract}

\section{Introduction}
\label{intro} The aspirations of 5G network are reflected in the white papers published by the leading telecommunication companies \cite{Huawei_Vision,5G_Radio_Access-Ericsson,NGMN_5G_White_Paper}. All of these white papers mention about the vision of more than $1$ Gbps data rate. To facilitate our discussion, we need to know what are the data that will be encrypted and decrypted in a 5G phone. We also need to know where and how many times the encryption and decryption will take place across the protocol stack on the phone. But 5G is not yet a reality and we do not have exact answers to these questions. So, we assume things about a 5G network and argue on the basis of those assumptions. We turn to the LTE (3GPP defined 4G network) network and different white papers \cite{5G_Nework_Architecture_Whitepaper,5G_Scenarios_and_Security_Design, Ericsson_white_paper_energy} published by the leading telecommunication companies to make the assumptions. In an LTE phone, the data that leave and enter the phone can be broadly classified into three categories. 
\begin{enumerate}
\item The control signals in between the phone and the core network
\item The control signals in between the phone and the radio network
\item The user data sent and received at the phone's application layer
\end{enumerate}
Both of the first two categories are confidentiality and integrity protected. For the third category, only the privacy is protected. Also note that, from the volume point of view, the major share of data belong to the third category. Comparing to the the third category, the cryptographic computation required for the data of first and second categories is negligible. The user data in an LTE phone is only once encrypted and decrypted across the protocol stack in PDCP layer. For a 5G phone, we assume the following:
\begin{enumerate}
\item User data will remain as the major share of the total data leaving and entering the phone.
\item The cryptographic computational need for the total volume of control signals will be negligible in comparison with that of the user data.
\item The user data will only once be encrypted and decrypted somewhere across the protocol stack.
\item In order to have a pessimistic estimation of energy consumption, we assume that integrity protection of user data will be introduced in 5G.
\end{enumerate}
Based on these assumptions, we will look into the cryptographic energy requirements and also the total energy requirements across the protocol stack of an LTE phone. Then we will scale up the data-rate from $100$ Mbps to $1$ Gbps and see how much extra pressure it puts on the battery of the phone in comparison with other energy hungry aspects of the phone like display and radio signalling.

The paper is organized by first giving a very short introduction to the architecture, the protocol stack and the cryptographic specifications of the LTE network in Section \ref{sec:lte_specifications}. In Section \ref{sec:throughput_and_energy_requirements_of_aes_snow3g}, we present the results found in the existing literature about the energy requirements of the two cryptosystems of interest, which are AES and SNOW 3G. In this section we also present the results about the energy consumption across the whole protocol stack of the link layer. In Section \ref{sec:overall_comparison} we present the energy consumption distribution of the whole phone among its different functional modules and show that the energy needed for cryptographic computation is not a threat for the battery life of the phone. 

\section{LTE Specifications}
\label{sec:lte_specifications}
An LTE network is comprised of broadly three components. The user equipment (UE), evolved radio network (E-UTRAN) known as radio network and evolved packet core (EPC) known as core network. The user equipment consists of a mobile equipment (ME) or a mobile phone for the context of this paper, and a universal integrated circuit card (UICC). The UICC hosts an application called subscriber identification module (SIM). In this paper when we refer to the user equipment, we mean it to be the mobile phone since the UICC does not have much functionality to consume a lot of energy.

\begin{figure}
\begin{center}
  \includegraphics[width=0.95\textwidth]{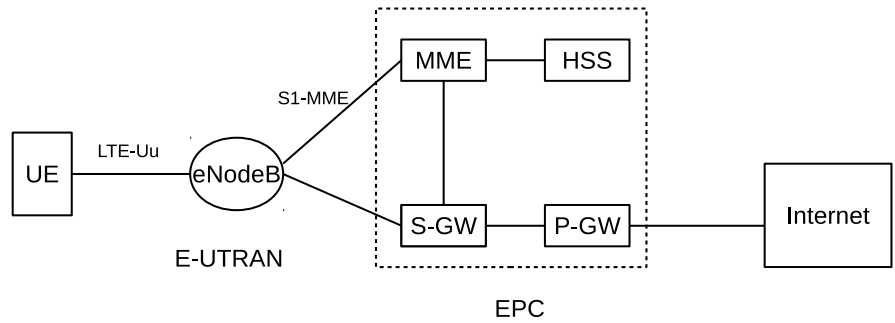}
\caption{LTE Architecture}
\label{fig:protocl_stack}       
\end{center}
\end{figure}

The UE is connected to the network via a radio link only with the radio network. The entity of the E-UTRAN that has the radio link with the UE is called eNodeB which is tradionally known as a base station. However, the UE also establishes a direct logical connection with an entity of the core network known as the mobility management entity (MME). This logical connection is used only for the control signals for the core network and hence we do not focus on it in this paper. The user data as mentioned in the introduction travels from the UE to the eNodeB. Figure \ref{fig:protocl_stack} shows the protocol stack that the data travel across at the UE and at eNodeB. The L1 layer is the physical layer. We logically bundle the packet data convergence protocol (PDCP), radio link control (RLC), and medium access control (MAC) layers as layer 2 (L2). All the encryption and decryption takes place at the PDCP layer \cite{3GPP_TS_36_323}.

\begin{figure}
\begin{center}
  \includegraphics[width=.98\textwidth]{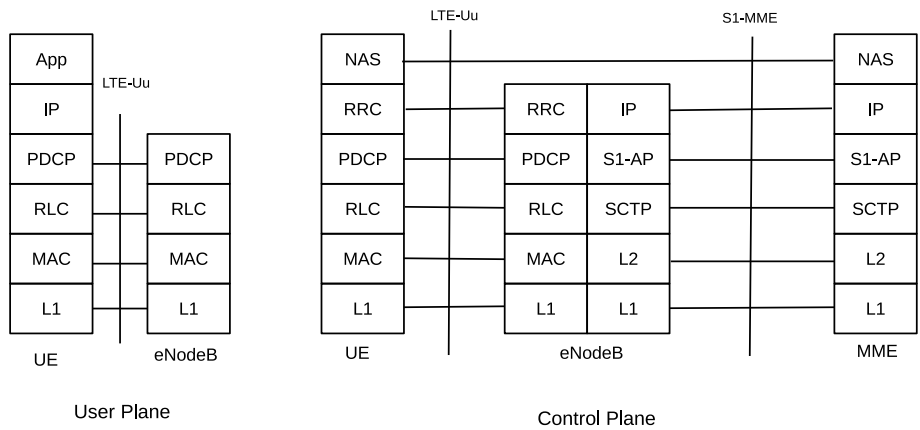}
\caption{Protocol stack in LTE Network}
\label{fig:protocl_stack}       
\end{center}
\end{figure}

According to \cite{3GPP_TS_33_401}, there are two mandatory sets of security algorithms in the 4th generation cellular network (LTE) developed by 3GPP. One is EEA1 in which the stream cipher SNOW 3G is used. The other is EEA2 in which the block cipher AES is used. The aspiration for the 5G networks is to obtain at least 1Gbps data-rate. The question arises if there are implementations of these two encryption systems that can achieve the required throughput and still be energy efficient enough to be used in a mobile phone.

\section{Throughput and Energy Requirements of AES and SNOW 3G}
\label{sec:throughput_and_energy_requirements_of_aes_snow3g}
In \cite{IIS_Ruhr_2009}, the authors showed in their experiment that the computing power of a single embedded processor even at high clock frequency is not enough to cope with the L2 requirements of LTE and next generation mobile devices. The AES decryption was identified as the major time critical software algorithm, demanding half of the execution time of entire L2 of downlink (DL). So, advanced hardware acceleration methods were required while keeping the energy and area requirements at a reasonable level for a mobile phone.

In a study conducted on the L2 DL \cite{IIS_Ruhr_2010}, the authors had shown that by a smart DMA (direct memory access) controller, the required throughput for LTE which is at most $100$ Mbps can be achieved. They used Faraday's $90$ nm CMOS technology, $128$-bit data path and $11$ round transformations for AES. However, to achieve this required throughput, the implementation consumed 9.5 mW of power whereas AES and SNOW 3G each required $.5$ and $.57$ mW of power respectively. So, the decryption consumes around 5 percent of the power budget of L2 DL. From energy point of view, it consumes $5.7$ mJ of energy to decrypt $1$ Gb of data. In \cite[Figure 6]{IIS_Ruhr_2010} the detailed comparison is presented. In Section \ref{sec:overall_comparison}, we will see that this is indeed a very small amount of energy when compared with total amount of energy consumed by the phone while exchanging bulk amount of network data.

From the experience of LTE \cite[Fig 9]{Mobisys_2012}, we see that the energy requirements of radio interface technology (downlink) in LTE increases linearly as the data rate increases but with a small slope. On the other hand, energy requirements for encryption increases linearly as the data rate increases with slope $1$. 5G has unprecedented data rate of 1 Gbps. Consequently, even though it is evident in LTE that ciphering is not very expensive, it needs to be rigorously investigated to conclude that it will not be very expensive in 5G. In the following sub-sections we present some implementations of AES and SNOW 3G.

\subsection{AES}
\label{sec:aes}
Since the adoption of Rijndael as AES by NIST, there have been a number of hardware implementations of AES. It is understandable that throughput and energy consumption are not mutually exclusive. In the beginning, the focus was completely on achieving high throughput. Over the time the need for high throughput, yet energy efficient implementations became more pressing and studies concerning the energy consumption of the implementations became available.
\begin{table}
\begin{center}
\begin{tabular}{|p{0.14\textwidth-2\tabcolsep}
				|p{0.08\textwidth-2\tabcolsep}
				|p{0.14\textwidth-2\tabcolsep}
				|p{0.14\textwidth-2\tabcolsep}
				|p{0.14\textwidth-2\tabcolsep}
				|p{0.14\textwidth-2\tabcolsep}
				|p{0.14\textwidth-2\tabcolsep}|
				}
\hline
Year & Ref & Tech(nm CMOS)& TP (Gbps) & Gates(K) & Power (mW) & Energy (mJ/Gb) \\
\hline
2001 & \cite{IBM_Japan_2001} & 110 & $2.6$ &$21.3$  & $-$ & $-$ \\ \hline
2001 & \cite{IBM_Japan_2001} & 110 & $0.311$ &$5.4$  & $-$ & $-$ \\ \hline
2001 & \cite{IBM_India_IIT_2001} & $-$ & $0.24$ &$4$  & $-$ & $-$ \\ \hline
2006 & \cite{Taiwan_2006} & 180 & $0.570$ & $-$ &20.34 & 35.68 \\ \hline
2006 & \cite{Taiwan_2006} & 350 & $0.569$ & $-$ &192.5 & 338.3 \\ \hline
2007 & \cite{IIT_Kharagpur_2007} & 180 & $0.384$ &$21$ & $-$  & $-$ \\ \hline
2009 & \cite{IME_China_Tsinghua_Univerisity_2009} & 180 & $1.16$ &$19.47$ & $-$ & $-$ \\ \hline
2009 & \cite{Ruhr_2009} &90&$ 1.00$ &$38$ &0.78 & $ .78$ \\ \hline
2011 & \cite{Ruhr_2011} & $-$ & $0.114$ &$ - $ &.02 & .24 \\ \hline
2012 & \cite{Pune_2012} & 180 & $1.6$ &$58.445$ &22.85 & 14.28 \\ \hline
\end{tabular}
\end{center}
\caption{AES Implementations}
\label{table:aes_implementation}
\end{table}

From Table \ref{table:aes_implementation}, we find the implementations in \cite{Ruhr_2009, Ruhr_2011, Pune_2012} are potential candidates for using in 5G since they meet the required throughput of 5G and present their energy requirements which enable us to make a meaningful analysis. In \cite{Ruhr_2011}, the authors present an implementation called SAME that achieves $114$ Mbps throughput using $2$ cores of the processor of a mobile phone. The implementation is based on slicing and merging the bytes of several data blocks to exploit processor's architecture width for multi-block encryption. According to \cite[Figure 8]{Ruhr_2011} the implementation is scalable with a speed up factor of $0.9$; i.e., if the throughput achieved by $1$ core is $T$, then $n > 1$ cores provide throughput $0.9nT$. According to \cite[Figure 9]{Ruhr_2011}, the SAME implementation achieves $5.5$ Mbps/$\mu$J. Consequently, it spends $114/5.5$ Mbps/(Mbps/$\mu$J) $=20.72$ $\mu$J $=.02$ mJ to encrypt/decrypt $114$ Mb. Now, by scaling up by $12$ different $2$-cores, it will achieve the throughput of $0.114 \cdot 12 \cdot 0.9 \approx 1$ Gbps while it will spend $0.02\cdot12=0.24$ mJ. Even though this implementation comes up as the most energy efficient in Table \ref{table:aes_implementation}, it is still not practical choice in 5G because a mobile phone with 24 cores is a far fetched idea even for 5G. In \cite{Ruhr_2009}, the authors present an application specific integrated circuit (ASIC) implementation based on Faraday's $90$ nm CMOS technology. They do not provide the exact throughput but provide the time required for processing one byte and the power need to process at that rate. In LTE, to achieve $100$ Mbps, 100 Kb of data is required to be processed by $0.6$ ms. Similarly we assume that in 5G, to achieve $1$ Gbps data rate, $1$ Mb of data need to be processed in $0.6$ ms. Consequently processing time of around $4$ ns per byte is required to achieve $1$ Gbps. In \cite[Figure 8]{Ruhr_2009} it is claimed that using one AES core with $128$-bit data path, processing time of $4$ ns per byte can be achieved while consuming $0.78$ mJ of energy per second. In \cite[Figure 10]{Ruhr_2009}, it is claimed that that processing time of $4$ ns per byte can be achieved using $2$ AES core by using even less energy, $0.72$ mJ per second. This implementation appears to be a very good candidate for a 5G phone.  In \cite{Pune_2012}, the implementation achieves the required throughput without any need of scaling up but it spends almost $30$ times more energy than that of \cite{Ruhr_2009}.

\subsection{SNOW 3G}
\label{sub-sec:snow3gp}
It appears that there have not been as many hardware implementations of SNOW 3G as there have been of AES. It may be attributed to the reason that AES is much more widely used in different protocols. As mentioned in Section \ref{sec:throughput_and_energy_requirements_of_aes_snow3g}, the implementation in \cite{IIS_Ruhr_2010} focuses on the throughput and energy efficiency of the whole L2 layer of LTE and doesn't give any account of the scalability of the SNOW 3G ciphering unit that can be used for much higher data rates than LTE. In \cite{IEEE_ICCT_2010}, the authors present a parallel implementation of SNOW 3G by exploiting the multi-core processor of a smart phone that can provide the required throughput of LTE. The authors used voltage and frequency scaling (VFS) to reduce the energy consumption. It achieves the energy efficiency of $22$ Mbps/$\mu$J while providing throughput of $100$ Mbps. So, it consumes $100/22$ Mbps/(Mbps/$\mu$J) $ = 4.5$ $\mu$J per second to achieve the throughput. If the technology was scalable to achieve the required throughput of 5G with a reasonable speed up factor, it would be an energy efficient solution. But the implementation depends on the cores of the processor of the phone to achieve the parallelism, and it seems it would take at least $10$ times more cores than that of the phone used in the original implementation. As a result we do not find it as an appealing implementation for a 5G phone. There has been an ASIC implementation by Elliptic Semiconductor Inc. that achieves $2.5$ Gbps throughput at $100$ Mhz frequency and $15$K gates as cited in \cite{Greece_SNOW3G}. In \cite{Greece_SNOW3G} the authors presented an ASIC implementation of SNOW 3G using the $130$ nm CMOS library with $1.2$ V core voltage and $25$K gates. At $249$ MHz they have been able to harness a throughput of $7.9$ Gbps. Though both of the implementations provide much more than the throughput required in 5G, we can't argue anything with them as no concrete power figures are found. In \cite{IP_cores}, IP Cores Inc. presents two implementations of SNOW 3G called SNOW3G1 as shown in table \ref{table:snow3g_implementation}.
\begin{table}
\begin{center}
\begin{tabular}{|p{0.22\textwidth-2\tabcolsep}
				|p{0.2\textwidth-2\tabcolsep}
				|p{0.2\textwidth-2\tabcolsep}
				|p{0.2\textwidth-2\tabcolsep}|				
				}
\hline
Technology & Max Frequency & Area/Resources & Throughput \\
\hline
TSMC 65 nm G+ & 302 MHz & 7,475 gates & 2.4 Gbps \\ \hline
TSMC 65 nm G+ & 943 MHz & 8,964 gates & 7.5 Gbps \\ \hline
\end{tabular}
\end{center}
\caption{SNOW3G1 in \cite{IP_cores}}
\label{table:snow3g_implementation}
\end{table}
They too do not provide any power/energy figures. Fortunately, in \cite{kolkata}, the authors have used the SNOW3G implementation of IP Cores Inc. and have estimated that using 4 parallel blocks of SNOW3G1 with hard macro storage, a throughput of 30 Gbps is achievable at $1650$ MHz while consuming $14.41$ mJ of energy per second. We scale down the frequency by $30$ times and expect the energy consumption per second will also be scaled down at the same proportion. According to that assumption, at $1650/30$ MHz $=55$ MHz, we should be able to harness the throughput of $1$ Gbps by spending $14.41/30$ mJ $=0.48$ mJ of energy. The authors estimated the power consumption on a gate-level netlist by back-annotating the switching activity and using Synopsys Power Compiler tool.

\section{Overall Comparison}
\label{sec:overall_comparison}
The overall energy consumption of a phone depends on the usage pattern of the user of the phone. Radio activities of the cellular network, lighting up the screen, touch screen and CPU are the commonly considered as the most energy hungry aspects of a smart phone \cite{Usenix_2010}. 

There are times when a smart phone remains idle and does nothing for a long duration of time. During this time it switches to a suspended state by transferring the state of the phone to the RAM. In suspended state the phone draws a minimal amount of energy from the battery to maintain the state in the memory and receive very limited control signals from the network to be able to receive the incoming traffic. In \cite{Usenix_2010}, the authors conducted an experiment on a $2.5$G phone and two cutting edge $3$G phones of the time. They showed that in suspended state, a $2.5$G phone drew $103$ mJ of energy per second whereas the 3G phones drew around $25$ mJ per second. There is another state when the phone is awake but no application is running. This state is called the idle state. In \cite{Usenix_2010}, the authors showed on the same phones that during idle state the amount of energy drawn was less than $350$ mJ per second. 

Normally, the time duration a smart phone is in a suspended or idle state is much longer than that of when it remains active. So, the energy consumption of the phone during idle or suspended state is very critical for the battery life. However, during these times, the phone hardly encrypts or decrypts any data except the control signals which are mostly paging messages. The reason is that the attach procedure takes place only when the user switches on the phone and tracking area update takes place frequently only when the user is travelling on a vehicle. However, even though paging itself is a burden for the phone from energy point of view, the cryptographic energy requirement for paging message is insignificant. According to \cite{Nokia_2013}, even with traditional paging mechanism, there are $1000$ paging messages for a phone in an hour, which is less than $1$ in a second. According to \cite{3GPP_TS_36_331}, the paging message is no longer than hundreds of bytes. The energy requirements for AES for this tiny amount of data is very insignificant to the total need $25$ mJ per second during the suspended state and of 300 mJ during the idle state.

To understand the energy expense of encryption, we need to focus on the total energy expense of the phone during the active states of the phone when encryption is also being performed. Such active states are phone call, web browsing, email, network data exchange (upload/download) and so on. We choose to focus on the case of network data exchange to argue our case. We assume that the phone would utilize its full download or upload capacity from the data volume point of view during the exchange. We will investigate this case for 2.5G, 3G and 4G phones to see the evolution the energy requirements.

In \cite{Usenix_2010}, the authors showed that the $2.5$G phone consumed around $700$ mJ of energy per second during the network data exchange. Around $640$ mJ of this energy budget is spend for cellular network activities. We know in $2.5$G, the maximum data rate can be 115 Kbps. At that rate the AES implementation in \cite{Ruhr_2009} would spend around $0.000078$ mJ of energy per second which is of course very insignificant. The authors of \cite{Usenix_2010} also showed that the 3G phones consumed a similar amount of total energy during data upload/download which is around $900$ mJ per second. Considering the connection exchanged the data at its full capacity (7.2 Mbps), the energy share for encryption is around $0.005$ mJ which is also very insignificant. 

Both in the 2.5G and 3G phone the major energy share for network data exchange is attributed to the radio transmission. However, there has been a significant change in the LTE radio technology and has become even more expensive from energy consumption point of view. In LTE, there are different radio states and the phone promotes and demotes to different states to save energy. As a result even though LTE becomes less energy efficient than 3G for small data transfer, it remains as efficient as 3G in large size data transfer. Also, there is a significant difference in the energy consumption of LTE uplink and downlink. According to \cite[Fig 9]{Mobisys_2012}, the LTE uplink consumes $3.2$ J of energy per second while uploading at the rate of $5$ Mbps. From the figure it appears that the energy consumption increases linearly with the uploading data rate with a factor of more than $1$. Downloading on the other hand, is less energy expensive, consuming $2.1$ J of energy per second at the rate of $19$ Mbps. The energy consumption while downloading also increases almost linearly but with a very small factor after $10$ Mbps. With screen off, the authors claimed that the energy was mostly consumed by the radio interfaces. The AES implementation in \cite{Ruhr_2011} consumes $.78$ mJ of energy per second providing throughput of $1$ Gbps. In order to come up with a loose bound, let us consider that the LTE uplink and downlink would consume the same amount of energy even when the data rate is at the theoretical peak, which is $100$ Mbps and around $90$ Mbps downlink and uplink respectively. Then the energy share of encryption is still bounded by $.04$ percent. It should be noted here that the high energy requirements in LTE are mostly attributed to its radio interface technology. Nevertheless, the radio technology will be different in 5G than that of LTE. Let us consider that the LTE draws $E_{lte}$ mJ of energy per second while transferring data at the theoretical maximum data rate. Let's assume that in 5G, the radio interface will draw $E_{lte}/a$ mJ of energy while providing the throughput of $1$ Gpps by using its new efficient radio technology. We know implementations of AES and SNOW 3G that take $.78$ and $.48$ mJ of energy per second to provide throughput of $1$ Gbps. So, the energy share of encryption in 5G is $\frac{.78a}{E_{lte}}\times 100 = .04a$ percent for AES and $.03a$ percent for SNOW 3G. Considering that the integrity protection will also be incorporated for user data, the cryptographic effort will at most be doubled and hence they will be at most $.08a$ and $.06a$ percent for AES and SNOW 3G respectively. 

However, we don't know the value of $a$ for certain. Energy efficiency is a major concern for 5G network. Operators explicitly mention a reduction of total network energy consumption by 50 percent despite an expected 1,000-fold traffic increase \cite{NGMN_Whie_Paper}. In \cite{Ericsson_white_paper_energy}, it concludes that 5G systems with high energy performance should be built on two design principles. One is to only be active and transmit when needed and the other is to only be active and transmit where needed. In our above discussion the $E_{lte}$ is considered as the energy only when data is being transmitted or received and doesn't include the energy when no data is being exchanged. We haven't found any account on how much more energy efficient 5G uplink and downlink will be. It is difficult to answer as there will be different radio interfaces involved. Search for a reasonably accepted value of $a$ would be a further research question. Very optimistically even we consider the value of $a$ to be $10$, the cryptographic energy share still remains below $1$ percent.

\section{Conclusion}
\label{sec:conclusion}
Number of energy efficient implementations of AES and SNOW 3G have been presented. Some of the implementations use the multiple cores of the CPU of the mobile phone while others use ASIC. We have found that the ASIC implementations can provide the required throughput of 5G. We made assumptions about a 5G network. Based on the assumptions and energy consumption related facts of LTE network available in the literature, we have shown that energy consumption for cryptographic computation is insignificant compared to the total energy need of the phone when bulk data transfer takes place. It should be noted that there might have other implementations of AES and SNOW 3G which are more energy efficient and provide the required throughput. But we did not look into any other implementations as it is evident that even with the implementations presented in this paper keeps the cryptographic energy share very low. However, as 3GPP advances on defining the 5G standards and more results on the energy consumption of the radio interfaces of 5G are published, the exact cryptographic energy share will be clearer.
\section{Acknowledgement}
\label{sec:acknowledgement}
Thanks to Kimmo J{\"a}rvinen for his help to understand the ASIC implementations and to Jarno Alanko for proofreading.

\end{document}